\documentclass[twocolumn,preprintnumbers,amsmath,amssymb]{revtex4}

\usepackage[dvips]{graphicx}
\usepackage{dcolumn}
\usepackage{bm}
\usepackage{epsfig}

\begin{document}


\title{Stray-field-induced modification of coherent spin dynamics}

\author{L. Meier\footnote[1]{Also at: Solid State Physics Laboratory, ETH Zurich, 8093
Z\"urich, Switzerland} and G. Salis} \affiliation{IBM Research,
Zurich Research Laboratory, S\"aumerstrasse 4, 8803 R\"uschlikon,
Switzerland}
\author{C. Ellenberger and K. Ensslin}
\affiliation{Solid State Physics Laboratory, ETH Zurich, 8093
Z\"urich, Switzerland}

\author{E. Gini}
\affiliation{FIRST Center for Micro- and Nanosciences, ETH Zurich,
8093 Zurich, Switzerland}

\date{December 21, 2005}

\begin{abstract}
Electron spins in an InGaAs semiconductor quantum well are used as a
magnetometer of magnetic stray-fields from patterned Fe stripes.
Using time-resolved Faraday rotation, the coherent precession of
quantum-well spins in the inhomogeneous field below the Fe stripes
is measured for varying magnetic fields. Comparing with reference
stripes made of Au, we find an enhancement of the spin precession
frequency proportional to the Fe magnetization, in line with a
decrease of the spin decay time, which is attributed to the
inhomogeneous magnetic stray-field in the quantum well layer.
\end{abstract}

\maketitle

The proposal of a spin-based analogue to an electro-optical
modulator, where spin-precession is induced in a semiconductor
channel by electric fields via spin-orbit
interaction~\cite{Datta1990}, has lead to intense research on
semiconductor spintronics, also driven by the prospects anticipated
by quantum computation relying on the electron
spin~\cite{Awschalom2002}. A central requirement for spin-based
information processing devices is the ability to coherently
manipulate spins. Besides spin-orbit effects, also
g-factor-engineered heterostructures~\cite{Jiang2001,Salis2001} or
magnetic stray-fields from patterned ferromagnetic
structures~\cite{Loss1998} are considered for spin manipulation. The
latter approach has the advantage of rather high magnetic fields
that are confined to small length scales. Different approaches have
been taken to characterize and detect magnetic stray-fields.
Magnetic-force microscopes~\cite{Martin1987} or scanning Hall
probes~\cite{Chang1992} provide spatially mapped field
distributions. The influence of stray-fields on nearby semiconductor
spin-states has been investigated by
photoluminescence~\cite{Schomig2004,Sakuma2004}, spin-flip light
scattering~\cite{Sakuma2003} and
cathodoluminescence~\cite{Kossut2001} in semiconductor quantum wells
(QWs). Since the Zeeman splitting in a QW is typically much smaller
than the photoluminescence linewidth, experiments have focused on
diluted magnetic semiconductors that exhibit a very large electron
g-factor and correspondingly a large Zeeman splitting. The spatially
varying Zeeman-splitting induced by a magnetic stray-field has also
been considered for spin-selective confinement of
electrons~\cite{Redlinski2005b}. Attempts to directly monitor the
influence of stray-fields on the dynamics of electron-spins have so
far remained elusive~\cite{Crowell1997}.

Here we demonstrate that the coherent spin-dynamics of QW electrons
are affected by magnetic stray-fields. An array of Fe stripes is
patterned on top of an InGaAs QW that is positioned close to the
semiconductor surface. The dynamics of the QW spins is monitored
using time-resolved Faraday rotation (TRFR)~\cite{Crooker1995}. When
the Fe stripes are magnetized perpendicular to the stripe length and
in-plane to the QW layer, the spin-precession frequency $\nu$
increases as compared to spins below a reference grating made of
non-magnetic Au. The increase is proportional to the magnetization
of the Fe grating, as measured by the magneto-optical Kerr effect
(MOKE). Varying the geometry of the stripes changes the
magnetization saturation field, which is reflected in the measured
spin dynamics. The decay time of the spin precession is reduced
below the ferromagnetic grating, due to the inhomogeneous
stray-field that leads to a dephasing of the spin ensemble. We
exclude that the observed effect stems from nuclear spins polarized
by optical orientation~\cite{Lampel1968} or imprinted from the
ferromagnetic stripes~\cite{Kawakami2001}. Micromagnetic simulations
confirm the order-of-magnitude of the stray-field, but give a
different dependence on the size of the stripes than the
measurements. This might be due to a non-ideal magnetization or due
to a non-uniform sampling of the spin distribution below the
grating.

We employ TRFR to track the coherent spin dynamics of QW electrons
in the time domain and obtain precise values of the electron spin
precession frequency $\nu$. The 40~nm thick InGaAs QW (8.8\% In) was
grown by Metal Organic Chemical Vapor Deposition on top of a GaAs
substrate and capped with 20~nm GaAs. Both well and cap are
$n$-doped with Si to ensure long spin lifetimes, the latter with a
$\delta$-doping in the middle of the layer, the former with a bulk
doping aimed at 5 $\times 10^{16} \textrm{cm}^{-3}$. On top of the
sample, Fe and Au gratings have been structured using electron beam
lithography and lift-off techniques. As an adhesion layer between
GaAs and the metal, we use 10~nm Ti. The gratings consist of
100~$\mu$m long bars with a thickness of 80nm. The Fe bars were
capped with 10~nm Al to prevent oxidation. We vary the width of the
bars as well as their spacing. The ratio between bar and gap width
is always 1. We have fabricated bars with widths 3~$\mu$m, 2~$\mu$m,
1~$\mu$m, and 0.5~$\mu$m and refer to the corresponding gratings as
3-3, 2-2, 1-1 and 0.5-0.5, meaning (bar width)-(gap width).

TRFR is measured in the Voigt geometry with an external magnetic
field $B_{ext}$ applied along x (in the QW plane and perpendicular
to the long axis of the bars). Electron spins are polarized along z
(perpendicular to the QW) by a circularly polarized pump pulse. The
helicity of the circular polarization is modulated with a
photo-elastic modulator at a frequency of 50~kHz, allowing the use
of lock-in amplifiers. We measure the Faraday rotation
$\Theta(\Delta t)$ of a linearly polarized probe pulse that is
delayed by a time $\Delta t$ with respect to the pump pulse. The
2~ps pulses of a mode-locked Ti:saphire laser are tuned to the
absorption edge of the QW at 870~nm and focused to a spot of about
15~$\mu$m in diameter. The pump (probe) beam has a power of 500
(60)~$\mu$W. The Faraday rotation $\Theta(\Delta t)$ is proportional
to the electron spin polarization along the probe beam direction and
can be expressed as
\begin{equation}\label{eq:trfr}
    \Theta(\Delta t) = \Theta_0 \cos{ \left( 2\pi \nu \Delta t \right)}
    \exp{(-\Delta t/T_2^*)},
\end{equation}
where $T_2^*$ is the spin lifetime of the ensemble and $\nu = g
\mu_B B_{tot}/h$ is given by the g-factor $g$ of QW electrons, the
Bohr magneton $\mu_B$ and the total local magnetic field $B_{tot}$
that includes $B_{ext}$, the magnetic stray-field of the
ferromagnetic grating $B_{s}$, and an effective magnetic field
$B_{n}$ resulting from hyperfine interaction with polarized nuclear
spins. Fitting Eq.~\ref{eq:trfr} to measurements of $\Theta(\Delta
t)$ yields $\nu$, from which $B_{tot}$ can be determined (provided
$g$ is known).

\begin{figure}
\includegraphics[width=80mm]{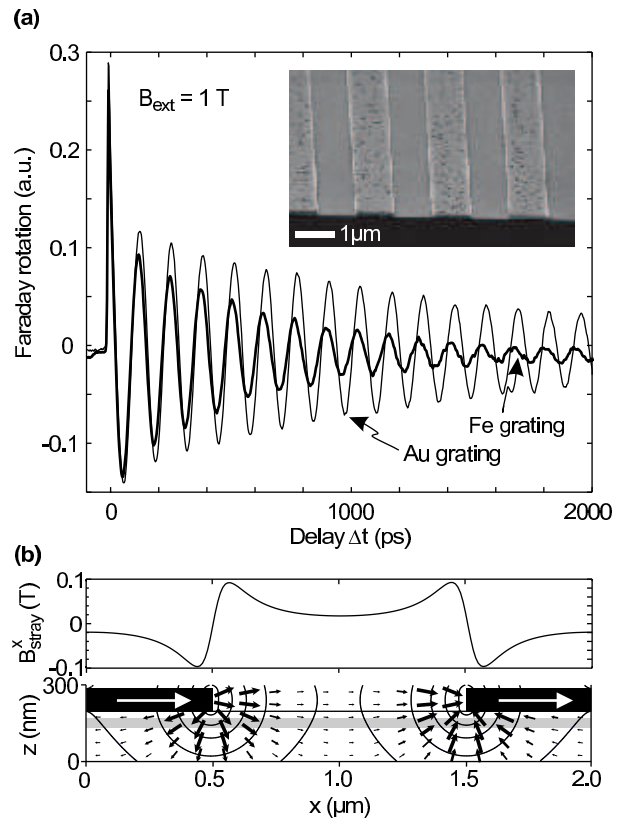}
\caption{\label{fig:fig1} \textbf{(a)} Faraday rotation measured at
40~K on a Fe and a Au grating with the dimensions 1-1. Due to the
magnetic stray-field of the Fe grating, electrons precess faster and
their lifetime is reduced. Inset: SEM-picture of a 1-1 Au grating.
\textbf{(b)} Calculated magnetic stray-field of two magnetized Fe
bars: x-component (top), direction and magnitude (bottom). Each bar
is 1$\mu$m wide and 80~nm thick. Lines indicate constant magnetic
fields of 500~mT, 200~mT, 100~mT, 50~mT and 20~mT (moving away from
a bar). The shaded area shows the location of the QW. }
\end{figure}

A numerical simulation of the magnetic stray-field of two Fe bars at
$B_{ext} = 1$~T obtained with the micromagnetic simulation tool
OOMMF~\cite{OOMMF} is shown in Fig. \ref{fig:fig1}(b). At $B_{ext} =
0$, they are magnetized along their long (easy) axis, and no
stray-field is expected along x. As $B_{ext}$ is increased, the bars
are magnetized in x-direction. Magnetization saturates at $B_{ext}$
of 100-400~mT, depending on the bar width. The stray-field in the QW
is strongly inhomogeneous and decays from 100~mT close to the bar
edge to values below 20~mT just 500~nm away from the edge.

Fits to the experimental data shown in Fig. \ref{fig:fig1}(a) yield
precession frequencies of $\nu_{Fe} = 7.82$~GHz and $\nu_{Au} =
7.68$~GHz on the Fe and Au grating, respectively. As there is no
magnetic stray-field originating from the Au grating and provided
$B_{n} = 0$ (see below), we can determine the stray-field averaged
over the gap width from $\langle B_{s} \rangle = h(\nu_{Fe} -
\nu_{Au})/g\mu_B = h \Delta \nu/g \mu_B$ to be 21~mT. The same fits
also yield $T_2^*=1510$~ps in the QW below the Au grating and 590~ps
below the Fe grating, which is reduced due to averaging effects in
the inhomogeneous stray-field.

\begin{figure}
\includegraphics[width=80mm]{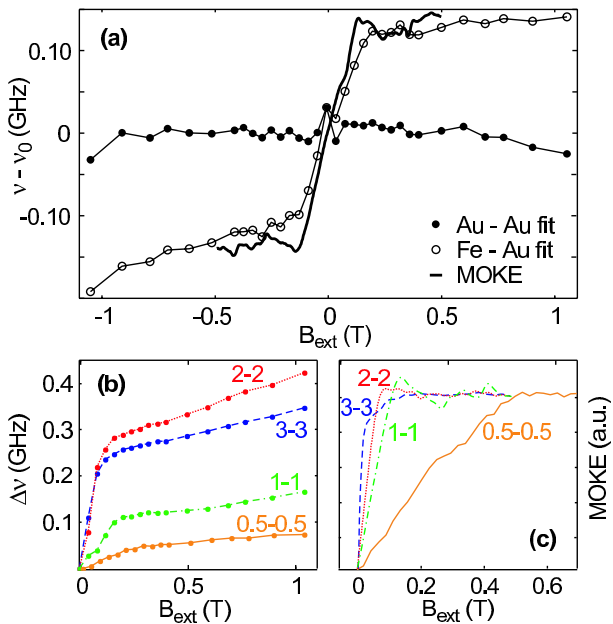}
\caption{\label{fig:fig2} (Color online) \textbf{(a)} Electron
precession frequency $\nu$ in the QW below the 1-1 Au grating (dots)
and below the 1-1 Fe grating (open circles) as a function of
$B_{ext}$. A linear fit to the Au data, $\nu_0(B_{ext})$, has been
subtracted. Electron precession in negative $B_{ext}$ is attributed
a negative $\nu$. Bold line: MOKE measurement of the Fe bar
magnetization, scaled to fit the y-scale. \textbf{(b)} Precession
frequency difference $\Delta \nu = \nu_{Fe} - \nu_{Au}$ for all
geometries ($T$ = 40~K). \textbf{(c)} MOKE measurements for all
geometries ($T$ = 295~K, background removed and normalized).}
\end{figure}

A more systematic study of the dependence on $B_{ext}$ is presented
in Fig. \ref{fig:fig2}(a), showing $\nu_{Fe/Au} - \nu_0$, with
$\nu_0 = g \mu_B B_{ext}/h$, where $g = 0.520$ has been determined
from a fit to $\nu_{Au}$. While $\nu_{Au}$ is clearly linear in
$B_{ext}$, the electrons precess faster below the Fe grating by an
amount that is proportional to the bar magnetization. The latter has
been measured independently with MOKE, using a continuous-wave laser
at a wavelength of 633~nm [bold line in Fig. \ref{fig:fig2}(a)]. We
suspect that the small linear increase in $\nu_{Fe}$ after
magnetization saturation is due to a slightly enhanced $g$-factor in
the QW below the Fe grating resulting from unequal strain exerted by
the Au and the Fe grating. This effect is more pronounced for wide
than for narrow bars, as visible in Fig. \ref{fig:fig2}(b).
Comparing the different geometries, we verify that a larger
$B_{ext}$ is required to fully magnetize the narrow bars compared to
the wide bars [Fig. \ref{fig:fig2}(b)], in agreement with
room-temperature MOKE measurements [Fig. \ref{fig:fig2}(c)].

We calculated the magnetization of the Fe bars and from this
$\textbf{B}_{s}(x,z)$. At $\Delta t = 0$ we uniformly distributed
electron spins in the QW between two Fe bars, all pointing along z.
For $\Delta t > 0$, the spins precess around a spatially varying
$\textbf{B}_{tot}(x,z) = \textbf{B}_{ext} + \textbf{B}_{s}(x,z)$.
The averaged z-component of these spins was calculated as a function
of $\Delta t$ and fitted with Eq. \ref{eq:trfr} to provide
$\nu_{Fe}$. From this simulation, we expect $\Delta \nu$ to be
between 0.05 and 0.5~GHz depending on the size of the gap between
the bars. Larger gaps exhibit a lower $\Delta \nu$ than small gaps.
For one individual Fe bar, $\textbf{B}_{s}$ does not depend on the
bar width, as it relies on the divergence of the magnetization,
which only depends on the boundaries of the bar. In the gap,
$\textbf{B}_{s}$ decays quickly, and thus for larger gaps the
averaged $\Delta \nu$ decreases.

However, we experimentally find the value of $\Delta \nu$ to depend
non-monotonically on the gap-size, as visible from Fig.
\ref{fig:fig2}(b). This was reproduced on two additional samples,
and could be explained by an additional field due to nuclear
polarization, inhomogeneous averaging over the gap due to
optical/electronic effects or non-perfect magnetization of the Fe
grating.

\begin{figure}
\includegraphics[width=80mm]{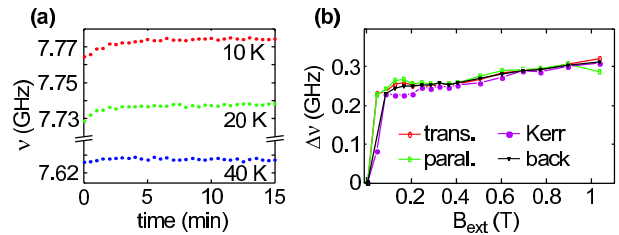}
\caption{\label{fig:fig3} (Color online) \textbf{(a)} Temperature
dependence of nuclear polarization at $B_{ext} = 1$~T. \textbf{(b)}
Measurement of $\Delta \nu$ on the 3-3 grating with different
set-ups: parallel and transverse probe polarization, Kerr geometry
and illuminating the sample from its backside.}
\end{figure}

In order to exclude nuclear contributions to the measured precession
frequency, we study how $\nu$ changes with time. After the sample
has been in the dark for at least 10 minutes, fast scans of
$\Theta(\Delta t)$ were performed. Electron precession becomes
faster on a time scale of minutes, which we attribute to an
establishing nuclear polarization~\cite{Kikkawa2000}. The amplitude
of this effect can be almost zeroed by increasing the temperature to
40~K [Fig. \ref{fig:fig3}(a)]. In order to minimize effects of
nuclear polarization, we performed all measurements shown here at a
temperature of 40~K. Since the photon helicity is modulated at
50~kHz, the effect of dynamic nuclear polarization is suppressed as
compared to optical pumping with constant
helicity~\cite{Salis2001b}.

We can further exclude effects of ferromagnetic imprinting
~\cite{Kawakami2001} to account for the observed difference: a
transparent, 7~nm thick Fe film evaporated on our sample with the
same technique as used for the gratings did not affect $\nu$ in the
QW (not shown). Also, we have measured curves similar to the one in
Fig. \ref{fig:fig2}(a) at temperatures between 10~K and 80~K and
found a maximum relative variation of $\Delta \nu$ of 10\%.

In the simulation, the stray-field is sampled uniformly in the gap
between the Fe bars. In the experiment, diffraction may lead to a
non-uniform excitation of spin-polarized electrons in the gap. Also
in-plane electric fields originating from the strain or contact
potential of the metallic bars might lead to a lateral
redistribution of the electrons on a picosecond time scale.
Furthermore the polarization rotation of the transmitted probe beam
weighs this electron spin distribution in a way that is affected by
near-field optical effects including surface plasmons. These
deviations are most important at the boundary of the Fe bars where
the stray-field is strongly non-uniform, which could explain the
discrepancy between the experimental results and the simplified
simulation.

In order to investigate the role of diffraction, we measured $\nu$
in different geometries and polarization configurations [Fig
\ref{fig:fig3}(b)]. Due to the metallic bars, diffraction should
depend on the probe beam polarization. However, no difference was
observed for the probe beam polarized perpendicular or parallel to
the bars. Surprisingly, also TRFR from the back (where the beams
first pass the QW before they are diffracted by the grating) give
similar results. This indicates that the equilibrium distribution of
electron spins in the QW is not strongly affected by the intensity
distribution of the pump beam. Measurements in Kerr geometry (where
the reflected probe beam is analyzed) also yield the same
$\nu(B_{ext})$. Further investigation is needed to understand the
role of diffraction in the weighting of the measured spins. Also,
imperfect magnetization of the Fe bars, e.g. induced by edge
roughness, might lead to discrepancies with our simulations.

In conclusion, we have observed an increase $\Delta\nu$ of the spin
precession frequency and a decrease of the decay time of QW
electrons below a ferromagnetic grating magnetized in an external
magnetic field. $\Delta\nu$ is proportional to the magnetization of
the grating and is not affected by nuclear polarization. This is
explained by the inhomogeneous magnetic stray-field. The dependence
of $\Delta\nu$ on the geometry of the grating can not be explained
by a uniform averaging over the expected stray-field in the QW
between the bars. This indicates a non-uniform distribution of
electron spins and/or a non-uniform sampling of the spins due to
near-field optical effects. Such hybrid ferromagnetic-semiconductor
structures could be useful for spin manipulation in spintronic
devices.

We thank O.~Homan from ETH Zurich for polishing the back side of our
wafers, M.~Tschudy for the evaporation of Fe, P.-O.~Jubert for help
with OOMMF, and R.~Allenspach and N.~Moll for fruitful discussions.


\newpage

\begin{thebibliography}{19}
\expandafter\ifx\csname
natexlab\endcsname\relax\def\natexlab#1{#1}\fi
\expandafter\ifx\csname bibnamefont\endcsname\relax
  \def\bibnamefont#1{#1}\fi
\expandafter\ifx\csname bibfnamefont\endcsname\relax
  \def\bibfnamefont#1{#1}\fi
\expandafter\ifx\csname citenamefont\endcsname\relax
  \def\citenamefont#1{#1}\fi
\expandafter\ifx\csname url\endcsname\relax
  \def\url#1{\texttt{#1}}\fi
\expandafter\ifx\csname urlprefix\endcsname\relax\def\urlprefix{URL
}\fi \providecommand{\bibinfo}[2]{#2}
\providecommand{\eprint}[2][]{\url{#2}}

\bibitem[{\citenamefont{Datta and Das}(1990)}]{Datta1990}
\bibinfo{author}{\bibfnamefont{S.}~\bibnamefont{Datta}} \bibnamefont{and}
  \bibinfo{author}{\bibfnamefont{B.}~\bibnamefont{Das}},
  \bibinfo{journal}{Appl. Phys. Lett.} \textbf{\bibinfo{volume}{56}},
  \bibinfo{pages}{665} (\bibinfo{year}{1990}).

\bibitem[{\citenamefont{Awschalom et~al.}(2002)\citenamefont{Awschalom, Loss,
  and Samarth}}]{Awschalom2002}
\bibinfo{author}{\bibfnamefont{D.~D.} \bibnamefont{Awschalom}},
  \bibinfo{author}{\bibfnamefont{D.}~\bibnamefont{Loss}}, \bibnamefont{and}
  \bibinfo{author}{\bibfnamefont{N.}~\bibnamefont{Samarth}},
  \emph{\bibinfo{title}{Semiconductor spintronics and quantum computation}},
  Nanoscience and technology (\bibinfo{publisher}{Springer},
  \bibinfo{address}{Berlin} \bibinfo{year}{2002}).

\bibitem[{\citenamefont{Jiang and Yablonovitch}(2001)}]{Jiang2001}
\bibinfo{author}{\bibfnamefont{H.~W.} \bibnamefont{Jiang}} \bibnamefont{and}
  \bibinfo{author}{\bibfnamefont{E.}~\bibnamefont{Yablonovitch}},
  \bibinfo{journal}{Phys. Rev. B} \textbf{\bibinfo{volume}{64}},
  \bibinfo{pages}{R41307} (\bibinfo{year}{2001}).

\bibitem[{\citenamefont{Salis et~al.}(2001{\natexlab{a}})\citenamefont{Salis,
  Kato, Ensslin, Driscoll, Gossard, and Awschalom}}]{Salis2001}
\bibinfo{author}{\bibfnamefont{G.}~\bibnamefont{Salis}},
  \bibinfo{author}{\bibfnamefont{Y.}~\bibnamefont{Kato}},
  \bibinfo{author}{\bibfnamefont{K.}~\bibnamefont{Ensslin}},
  \bibinfo{author}{\bibfnamefont{D.~C.} \bibnamefont{Driscoll}},
  \bibinfo{author}{\bibfnamefont{A.~C.} \bibnamefont{Gossard}},
  \bibnamefont{and} \bibinfo{author}{\bibfnamefont{D.~D.}
  \bibnamefont{Awschalom}}, \bibinfo{journal}{Nature}
  \textbf{\bibinfo{volume}{414}}, \bibinfo{pages}{619}
  (\bibinfo{year}{2001}{\natexlab{a}}).

\bibitem[{\citenamefont{Loss and DiVincenzo}(1998)}]{Loss1998}
\bibinfo{author}{\bibfnamefont{D.}~\bibnamefont{Loss}} \bibnamefont{and}
  \bibinfo{author}{\bibfnamefont{D.~P.} \bibnamefont{DiVincenzo}},
  \bibinfo{journal}{Phys. Rev. A} \textbf{\bibinfo{volume}{57}},
  \bibinfo{pages}{120} (\bibinfo{year}{1998}).

\bibitem[{\citenamefont{Martin and Wickramasinghe}(1987)}]{Martin1987}
\bibinfo{author}{\bibfnamefont{Y.}~\bibnamefont{Martin}} \bibnamefont{and}
  \bibinfo{author}{\bibfnamefont{H.~K.} \bibnamefont{Wickramasinghe}},
  \bibinfo{journal}{Appl. Phys. Lett.} \textbf{\bibinfo{volume}{50}},
  \bibinfo{pages}{1455} (\bibinfo{year}{1987}).

\bibitem[{\citenamefont{Chang et~al.}(1992)\citenamefont{Chang, Hallen,
  Harriott, Hess, Kao, Kwo, Miller, Wolfe, van~der Ziel, and
  Chang}}]{Chang1992}
\bibinfo{author}{\bibfnamefont{A.~M.} \bibnamefont{Chang}},
  \bibinfo{author}{\bibfnamefont{H.~D.} \bibnamefont{Hallen}},
  \bibinfo{author}{\bibfnamefont{L.}~\bibnamefont{Harriott}},
  \bibinfo{author}{\bibfnamefont{H.~F.} \bibnamefont{Hess}},
  \bibinfo{author}{\bibfnamefont{H.~L.} \bibnamefont{Kao}},
  \bibinfo{author}{\bibfnamefont{J.}~\bibnamefont{Kwo}},
  \bibinfo{author}{\bibfnamefont{R.~E.} \bibnamefont{Miller}},
  \bibinfo{author}{\bibfnamefont{R.}~\bibnamefont{Wolfe}},
  \bibinfo{author}{\bibfnamefont{J.}~\bibnamefont{van~der Ziel}},
  \bibnamefont{and} \bibinfo{author}{\bibfnamefont{T.~Y.} \bibnamefont{Chang}},
  \bibinfo{journal}{Appl. Phys. Lett.} \textbf{\bibinfo{volume}{61}},
  \bibinfo{pages}{1974} (\bibinfo{year}{1992}).

\bibitem[{\citenamefont{Sch\"{o}mig et~al.}(2004)\citenamefont{Sch\"{o}mig,
  Forchel, Halm, Bacher, Puls, and Henneberger}}]{Schomig2004}
\bibinfo{author}{\bibfnamefont{H.}~\bibnamefont{Sch\"{o}mig}},
  \bibinfo{author}{\bibfnamefont{A.}~\bibnamefont{Forchel}},
  \bibinfo{author}{\bibfnamefont{S.}~\bibnamefont{Halm}},
  \bibinfo{author}{\bibfnamefont{G.}~\bibnamefont{Bacher}},
  \bibinfo{author}{\bibfnamefont{J.}~\bibnamefont{Puls}}, \bibnamefont{and}
  \bibinfo{author}{\bibfnamefont{F.}~\bibnamefont{Henneberger}},
  \bibinfo{journal}{Appl. Phys. Lett.} \textbf{\bibinfo{volume}{84}},
  \bibinfo{pages}{2826} (\bibinfo{year}{2004}).

\bibitem[{\citenamefont{Sakuma et~al.}(2004)\citenamefont{Sakuma, Hykomi,
  Souma, Murayama, and Oka}}]{Sakuma2004}
\bibinfo{author}{\bibfnamefont{M.}~\bibnamefont{Sakuma}},
  \bibinfo{author}{\bibfnamefont{K.}~\bibnamefont{Hykomi}},
  \bibinfo{author}{\bibfnamefont{I.}~\bibnamefont{Souma}},
  \bibinfo{author}{\bibfnamefont{A.}~\bibnamefont{Murayama}}, \bibnamefont{and}
  \bibinfo{author}{\bibfnamefont{Y.}~\bibnamefont{Oka}},
  \bibinfo{journal}{Appl. Phys. Lett.} \textbf{\bibinfo{volume}{85}},
  \bibinfo{pages}{6203} (\bibinfo{year}{2004}).

\bibitem[{\citenamefont{Sakuma et~al.}(2003)\citenamefont{Sakuma, Hyomi, Souma,
  Murayama, and Oka}}]{Sakuma2003}
\bibinfo{author}{\bibfnamefont{M.}~\bibnamefont{Sakuma}},
  \bibinfo{author}{\bibfnamefont{K.}~\bibnamefont{Hyomi}},
  \bibinfo{author}{\bibfnamefont{I.}~\bibnamefont{Souma}},
  \bibinfo{author}{\bibfnamefont{A.}~\bibnamefont{Murayama}}, \bibnamefont{and}
  \bibinfo{author}{\bibfnamefont{Y.}~\bibnamefont{Oka}}, \bibinfo{journal}{J.
  Appl. Phys.} \textbf{\bibinfo{volume}{94}}, \bibinfo{pages}{6423}
  (\bibinfo{year}{2003}).

\bibitem[{\citenamefont{Kossut et~al.}(2001)\citenamefont{Kossut, Yamakawa,
  Nakamura, Cywi\'{n}ski, Fronc, Czeczott, Wr\'{o}bel, Kyrychenko, Wojtowicz,
  and Takeyama}}]{Kossut2001}
\bibinfo{author}{\bibfnamefont{J.}~\bibnamefont{Kossut}},
  \bibinfo{author}{\bibfnamefont{I.}~\bibnamefont{Yamakawa}},
  \bibinfo{author}{\bibfnamefont{A.}~\bibnamefont{Nakamura}},
  \bibinfo{author}{\bibfnamefont{G.}~\bibnamefont{Cywi\'{n}ski}},
  \bibinfo{author}{\bibfnamefont{K.}~\bibnamefont{Fronc}},
  \bibinfo{author}{\bibfnamefont{M.}~\bibnamefont{Czeczott}},
  \bibinfo{author}{\bibfnamefont{J.}~\bibnamefont{Wr\'{o}bel}},
  \bibinfo{author}{\bibfnamefont{F.}~\bibnamefont{Kyrychenko}},
  \bibinfo{author}{\bibfnamefont{T.}~\bibnamefont{Wojtowicz}},
  \bibnamefont{and} \bibinfo{author}{\bibfnamefont{S.}~\bibnamefont{Takeyama}},
  \bibinfo{journal}{Appl. Phys. Lett.} \textbf{\bibinfo{volume}{79}},
  \bibinfo{pages}{1789} (\bibinfo{year}{2001}).

\bibitem[{\citenamefont{Redli\'{n}ski et~al.}(2005)\citenamefont{Redli\'{n}ski,
  Wojtowicz, Rappoport, Libal, Furdyna, and Jank\'{o}}}]{Redlinski2005b}
\bibinfo{author}{\bibfnamefont{P.}~\bibnamefont{Redli\'{n}ski}},
  \bibinfo{author}{\bibfnamefont{T.}~\bibnamefont{Wojtowicz}},
  \bibinfo{author}{\bibfnamefont{T.~G.} \bibnamefont{Rappoport}},
  \bibinfo{author}{\bibfnamefont{A.}~\bibnamefont{Libal}},
  \bibinfo{author}{\bibfnamefont{J.~K.} \bibnamefont{Furdyna}},
  \bibnamefont{and}
  \bibinfo{author}{\bibfnamefont{B.}~\bibnamefont{Jank\'{o}}},
  \bibinfo{journal}{Phys. Rev. B} \textbf{\bibinfo{volume}{72}},
  \bibinfo{pages}{085209} (\bibinfo{year}{2005}).

\bibitem[{\citenamefont{Crowell et~al.}(1997)\citenamefont{Crowell, Nikitin,
  Awschalom, Flack, Samarth, and Prinz}}]{Crowell1997}
\bibinfo{author}{\bibfnamefont{P.~A.} \bibnamefont{Crowell}},
  \bibinfo{author}{\bibfnamefont{V.}~\bibnamefont{Nikitin}},
  \bibinfo{author}{\bibfnamefont{D.~D.} \bibnamefont{Awschalom}},
  \bibinfo{author}{\bibfnamefont{F.}~\bibnamefont{Flack}},
  \bibinfo{author}{\bibfnamefont{N.}~\bibnamefont{Samarth}}, \bibnamefont{and}
  \bibinfo{author}{\bibfnamefont{G.~A.} \bibnamefont{Prinz}},
  \bibinfo{journal}{J. Appl. Phys.} \textbf{\bibinfo{volume}{81}},
  \bibinfo{pages}{5441} (\bibinfo{year}{1997}).

\bibitem[{\citenamefont{Crooker et~al.}(1995)\citenamefont{Crooker, Awschalom,
  and Samarth}}]{Crooker1995}
\bibinfo{author}{\bibfnamefont{S.~A.} \bibnamefont{Crooker}},
  \bibinfo{author}{\bibfnamefont{D.~D.} \bibnamefont{Awschalom}},
  \bibnamefont{and} \bibinfo{author}{\bibfnamefont{N.}~\bibnamefont{Samarth}},
  \bibinfo{journal}{Selected Topics in Quantum Electronics, IEEE Journal of}
  \textbf{\bibinfo{volume}{1}}, \bibinfo{pages}{1082} (\bibinfo{year}{1995}).

\bibitem[{\citenamefont{Lampel}(1968)}]{Lampel1968}
\bibinfo{author}{\bibfnamefont{G.}~\bibnamefont{Lampel}},
  \bibinfo{journal}{Phys. Rev. Lett.} p. \bibinfo{pages}{491}
  (\bibinfo{year}{1968}).

\bibitem[{\citenamefont{Kawakami et~al.}(2001)\citenamefont{Kawakami, Kato,
  Hanson, Malajovich, Stephens, Johnston-Halperin, Salis, Gossard, and
  Awschalom}}]{Kawakami2001}
\bibinfo{author}{\bibfnamefont{R.~K.} \bibnamefont{Kawakami}},
  \bibinfo{author}{\bibfnamefont{Y.}~\bibnamefont{Kato}},
  \bibinfo{author}{\bibfnamefont{M.}~\bibnamefont{Hanson}},
  \bibinfo{author}{\bibfnamefont{I.}~\bibnamefont{Malajovich}},
  \bibinfo{author}{\bibfnamefont{J.~M.} \bibnamefont{Stephens}},
  \bibinfo{author}{\bibfnamefont{E.}~\bibnamefont{Johnston-Halperin}},
  \bibinfo{author}{\bibfnamefont{G.}~\bibnamefont{Salis}},
  \bibinfo{author}{\bibfnamefont{A.~C.} \bibnamefont{Gossard}},
  \bibnamefont{and} \bibinfo{author}{\bibfnamefont{D.~D.}
  \bibnamefont{Awschalom}}, \bibinfo{journal}{Science}
  \textbf{\bibinfo{volume}{294}}, \bibinfo{pages}{131} (\bibinfo{year}{2001}).

\bibitem[{\citenamefont{Kikkawa and Awschalom}(2000)}]{Kikkawa2000}
\bibinfo{author}{\bibfnamefont{J.~M.} \bibnamefont{Kikkawa}} \bibnamefont{and}
  \bibinfo{author}{\bibfnamefont{D.~D.} \bibnamefont{Awschalom}},
  \bibinfo{journal}{Science} \textbf{\bibinfo{volume}{287}},
  \bibinfo{pages}{473} (\bibinfo{year}{2000}).

\bibitem[{\citenamefont{Salis et~al.}(2001{\natexlab{b}})\citenamefont{Salis,
  Awschalom, Ohno, and Ohno}}]{Salis2001b}
\bibinfo{author}{\bibfnamefont{G.}~\bibnamefont{Salis}},
  \bibinfo{author}{\bibfnamefont{D.~D.} \bibnamefont{Awschalom}},
  \bibinfo{author}{\bibfnamefont{Y.}~\bibnamefont{Ohno}}, \bibnamefont{and}
  \bibinfo{author}{\bibfnamefont{H.}~\bibnamefont{Ohno}},
  \bibinfo{journal}{Phys. Rev. B} \textbf{\bibinfo{volume}{64}},
  \bibinfo{eid}{195304} (\bibinfo{year}{2001}{\natexlab{b}}).


\bibitem{OOMMF} \texttt{http://math.nist.gov/oommf/}

\end{thebibliography}
\end{document}